\documentclass[preprint,amsmath,amssymb]{revtex4}

\usepackage{graphicx}
\usepackage{dcolumn}
\usepackage{bm}
\input{epsf.tex}
\def\ba{\begin{eqnarray}}
\def\ea{\end{eqnarray}}

\begin{document}

\def\inseps#1#2{\def\epsfsize##1##2{#2##1} \centerline{\epsfbox{#1}}}

\input{epsf.tex}
\input{psfig.tex}

\def\ba{\begin{eqnarray}}
\def\ea{\end{eqnarray}}
\def\be{\begin{equation}}
\def\ee{\end{equation}}
\def\tr{{\rm tr}}
                 
\def\redot{}
\def\releft{[}
\def\reright{]}

\title{Global Structure of the Colliding Bubble Braneworld Universe%
\footnote{
To appear in the Proceedings of JENAM 2002, Workshop on Extra Dimensions
and Varying Fundamental Constants.\break 
E-mail: J.J.Blanco-Pillado@damtp.cam.ac.uk, M.A.Bucher@damtp.cam.ac.uk}
} 

\author{Jos\'e J. Blanco-Pillado\thanks{E-mail: J.J.Blanco-Pillado@damtp.cam.ac.uk } 
and Martin Bucher
} 

\date{14 October 2002}

\begin{abstract}
The one-brane Randall-Sundrum model offers an example of a model with
an ``infinite" extra dimension in which ordinary gravity is recovered
at large distances and the usual (3+1)-dimensional cosmology 
at late cosmic times. This is possible because the ``bulk" has the geometry 
of anti de Sitter space, the curvature length $\ell $ of which delineates 
the (3+1)-dimensional behavior at large distances from the (4+1)-dimensional 
behavior at short distances. This spacetime, however, 
possesses a past Cauchy horizon
on which initial data must be specified in a natural and convincing way. 
A more complete story is required that singles out some set of initial
conditions to resolve the ``bulk" smoothness and horizon problems. One
such complete story is offered by the colliding bubble braneworld universe,
where bubbles filled with $AdS^5$ nucleate from $dS^5$ or $M^5$
through quantum tunnelling. A pair of such colliding bubbles forms a
 Randall-Sundrum-like universe in the future of the collision. Because of the 
symmetry of bubbles produced through quantum tunnelling, the resulting
universe is spatially homogeneous and isotropic at leading order, and the 
perturbations at the next order are completely well defined and calculable. 
In this contribution we discuss the possible global structure of such a 
spacetime. 

\end{abstract}

\maketitle

\section{Introduction}  
The Randall-Sundrum model \cite{rs} demonstrates how a braneworld cosmology
with an ``infinite" extra dimension may be achieved in such a way
that at large distances gravity behaves as ordinary (3+1) dimensional
gravity \cite{gt} and in the late time limit the ordinary FRW cosmology
is recovered \cite{langloisa}. This is achieved by postulating a bulk
 with the geometry of $AdS^5$ into which our brane is embedded. 
This scenario, however, is incomplete because of the usual 
Randall-Sundrum coordinates are bounded 
in the past by a Cauchy horizon. Some more complete story establishing 
well-defined initial conditions on this horizon is required. 

Unlike de Sitter space, anti de Sitter space lacks the ``no hair"
property. Perturbations in AdS forever retain their initial
amplitude. Therefore some sort of a beginning that resolves the bulk horizon 
and smoothness problems \cite{trodden} is required to provide a 
complete story. 

The colliding bubble braneworld universe \cite{bucher}, in which a 
(3+1)-brane surrounded 
by $AdS^5$ arises from the collisions of two bubbles filled with $AdS^5$ 
nucleating in $dS^5$, or $M~^5$, offers one such possible beginning. 

In this contribution we discuss some issues relating to the global structure of
the spacetime in the colliding bubble braneworld universe. It has been pointed
out that an isolated bubble filled with anti de Sitter space is likely to 
generate a spacelike singularity in its interior due to an instability that 
occurs during the collapse phase. We show how a collision may
partially avert the formation of such a singularity. The collisions of 
bubbles arising from quantum tunnelling were considered by Hawking, Moss, 
and Stewart 
\cite{hawkingb} and later in the thin-wall approximation with gravity
taken into account by Wu \cite{wu}. Some of the issues considered in this 
paper (e.g., Cauchy horizons, the consequence of a small perturbation 
on the global structure) parallel those arising 
in the determination of the global 
structure of a Reissner-Nordstrom black hole produced 
from a realistic collapse, as discussed for example in Poisson and 
Israel \cite{poisson}. 

\begin{figure}[h]
\vskip 2.4in
{\hskip -1in
\begin{picture}(600,150)
\put(168,115){(a)}
\put(310,115){(b)}
\put(168,0){(c)}
\put(310,0){(d)}

\put(125,20){\leavevmode\epsfxsize=3.5in\epsfbox{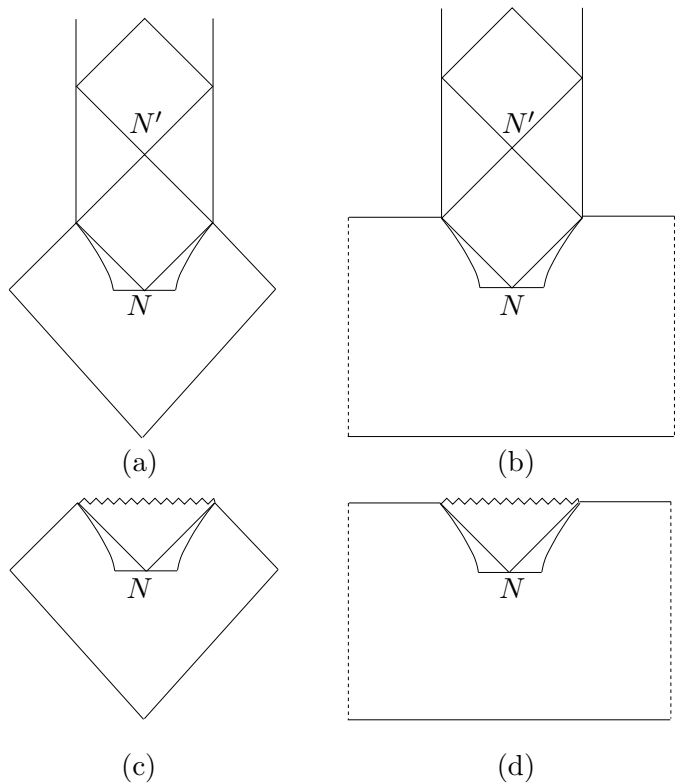}}

\put(170,174){$N$}
\put(171,243){$N'$}
\put(170,67){$N$}

\put(311,174){$N$}
\put(312,243){$N'$}

\put(311,67){$N$}

\end{picture}
}
\caption{\small\baselineskip=8pt  
{\bf Isolated bubbles filled with AdS space.}
For idealized thin-wall bubbles, in which a brane of
vanishing thickness and given tension mediates between
Minkowski space [panel (a)] or de Sitter space [panel (b)]
on the one hand and anti de Sitter space on
the other, the conformal diagrams showing the global
structure of the spacetimes are indicated. Curiously,
an initially globally hyperbolic spacetime evolves into
a spacetime that ceases to be so in the interior of the
bubble, because beyond a Cauchy horizon, which coincides
with the past lightcone of $N',$ additional boundary data
is required from the edge of the AdS infinite vertical
strip. However, for more realistic AdS bubbles considered
to lowest order in the semi-classical $(\hbar \to 0)$
expansion, the global structures are modified to those
shown in panels (c) and (d) where a spacelike singularity
forms owing to an instability of the perfect symmetric
solution.
}
\end{figure}

\begin{figure}[h]
\vskip .79in
{\hskip -0.9in
\begin{picture}(600,150)
\put(138,10){N}
\put(138,131){N'}
\put(70,65){$S_{\infty}$}
\put(191,40){C}
\put(137,76){$C_s$}
\put(200,105){$LB_{\infty}$}
\put(170,130){$H_c$}
\put(292,109){$H_c$}
\put(320,109){$H_c$}
\put(313,90){$LB_{\infty}$}
\put(260,80){$S_{L}$}
\put(355,80){$S_{R}$}
\put(307,46){$C$}
\put(60,0){\leavevmode\epsfxsize=4.8in\epsfbox{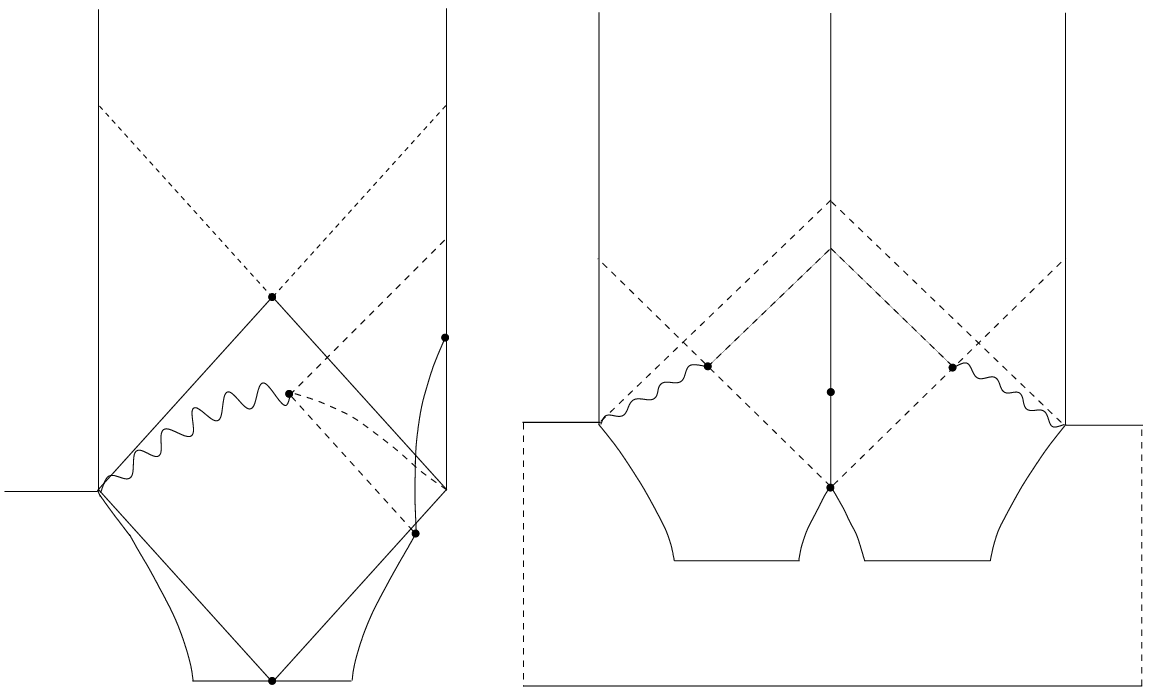}}
\end{picture}
}
\vskip -10pt
\caption{\small\baselineskip=8pt
{\small \bf Possible global structure of two colliding bubbles considered
to lowest order in the semi-classical $(\hbar \to 0)$ limit.}
A collision with a second bubble spoils the symmetry that
led to the perfect focusing of the scalar field.
It was this perfect focusing that ensured the formation of a
singularity. For two colliding bubbles, a shock wave of debris
emanates from the collision, perturbing the perfect AdS bulk.
If this perturbation is sufficiently strong, it is plausible that
a hole opens up into the would-be singularity, allowing the spacetime
to be extended into the full vertical AdS strip.
Because of causality, the singularity from $S_N$ to $C_S$
cannot be averted. However, the rest of the singularity
above the null curve from $C$ to $C_S$ can be avoided.
A Cauchy horizon $H_C$ emanates from $C_S$ upward along the diagonal.
This horizon indicates the boundary beyond which the initial value
problem is ill-defined without additional data from the AdS boundary
and from the ``other side of the singularity". The globally hyperbolic
spacetime bounded by the horizon, however, suffices to provide a
`complete story' for what happens on our brane, because this brane
propagates to the conformal boundary of the AdS strip, to $LB_\infty ,$
presumably before the intersection of $H_C$ with the conformal boundary.
Because of the divergence of the conformal factor on the boundary,
a infinite proper time elapses on our brane prior to $LB_\infty .$
Panel (b) shows the entire proposed conformal diagram for the two
colliding bubbles. The diagram includes two parallel AdS strips side by
side. These strips, however, are not connected to each other
because of the divergence of the conformal factor on the line
separating them running from $LB_\infty $ upward. From $C$ to
$LB_\infty ,$ the worldline of our local brane, the conformal
factor is finite and a $Z_2$ symmetry across the brane is present at
lowest order.
}
\end{figure}

\section{Interior Stability of AdS Bubbles}

In Coleman and de Luccia and later in Abbott and Coleman
\cite{ac}, it was pointed out that the interior of a bubble filled with
AdS is likely to be unstable toward the formation of a spacelike
singularity. Unlike most instabilities, which result from small
perturbations from a symmetric solution that progressively grow 
and eventually blow up, in this case the singularity arises 
rather from the absence, or the smallness, of perturbations from the 
symmetric model solution.

The origin of the instability is as follows. For tunnelling
described by a scalar field order parameter, the Euclidean 
instanton never takes one all the way to the true minimum, but rather at 
best very close but slightly displaced from there.
Said another way, even very thin-wall bubbles always have some tail 
of their wall that extends into the forward lightcone of the bubble
 nucleation center.
The evolution of the field in the lightcone interior is governed by
the equation $\ddot \phi +4(\dot a/a)\dot \phi =-V_{,\phi },$
where the derivatives are with respect to the proper time $\tau $ from
the nucleation center. While the bubble interior is expanding,
$(\dot a/a)$ is positive, and this term 
dampens any oscillations about the true minimum.
For a Minkowski or de Sitter interior, $(\dot a/a)$ is always positive.
However, when the bubble interior has the geometry of anti de Sitter
space, $a(\tau )=\ell \sin [\tau /\ell ]$ where $\ell $ is the AdS curvature
length, and for $\tau \ge (\pi \ell /2)$
the aforementioned dampening turns into 
anti-dampening during this collapse phase,
causing the oscillations to blow up near $\tau \approx \pi \ell $ 
(except for the
implausible case where the phase of the oscillations in finely tuned
with infinite precision by proper choice of the potential). So far
we have ignored gravitation backreaction, but the gravity of the scalar field
only hastens the formation of a singularity, turning what would simply be
a divergence in the energy density into a spacetime singularity, of the same
sort as the initial singularity of a hyperbolic FRW universe with the arrow
of time reversed. 

The spacetime singularity results because the Euclidean instanton has 
finely-tuned the wave front of the bubble wall tail, aiming it exactly 
toward the  
antipodal point of the nucleation center $N.$ There is nothing mysterious
about this singularity. The AdS space inside the bubble simply acts as a sort
of perfect lens, with absolutely no aberration and of infinite size so that
diffraction does not cut off the energy density at the focus. In the language
of geometric optics, the `rays' of the scalar field, which here are the 
timelike geodesics emanating from the nucleation center $N,$ are re-focused at 
the conjugate point $N'.$ A scalar field
in flat space with similarly perfectly focused initial conditions, as 
contrived as they may be, would behave exactly the same.

Having explained the nature of the singularity in the symmetric solution,
we now consider how it might be avoided. The obvious way is to spoil the 
perfect focusing. As Abbott and Coleman pointed out, in this instance
the Hawking singularity theorems do not pose an obstacle to avoiding the 
singularity altogether,
because the relevant singularity theorem demonstrates that a singularity must
arise if the spacetime is globally hyperbolic. However, a thin-wall solution
of a bubble nucleating from Minkowski  
or $dS$ space and tunnelling to $AdS$, whose 
spacetime conformal diagram is shown in Fig.~1, is not globally hyperbolic.
Consequently, the singularity theorems cannot be applied. The discussion
above assumed the lowest order of the $\hbar \to 0$ semi-classical 
limit. What happens at finite $\hbar$ remains an open question.

Of primordial importance for the colliding bubble model is understanding
the consequences of the perturbation presented by the collisions of two
AdS bubbles. In the idealized case, as considered in 
refs.~\cite{bucher,capo,gtb}, upon colliding, the bubbles deposit all their 
excess energy (beyond that required to form in its unexcited state
the intermediate brane on which we live)
on the brane in the form of radiation-matter. 
This idealization of no energy escaping into the bulk
is a caricature, just as that of an infinitely thin
bubble wall with no tail. There will always be some, if not a lot of leakage
from the collision into the bulk, which spoils the perfect focusing
symmetry in the causal future of the collision surface.
It is highly
plausible that the perturbation from this wave into the bulk
disrupts the formation of the singularity. 

The previous analysis tacitly presupposes that a bubble wall, to the extent
that it is not perfectly thin-wall, can be represented as a kink
in a scalar order-parameter field. It is not entirely evident that the
``tail" of a ``brane" bubble wall would behave in the same way. However,
we expect that branes, even if they are ``fundamental," are dressed with
some sort of tail similar to that of a scalar field kink because of couplings
to other fields and radiative corrections. 

To determine the global structure of the spacetime resulting from a
realistic collision requires numerical simulations, which are currently
in progress \cite{capob}. It is nevertheless interesting to speculate
on the possible outcome. In Fig. 2 we indicate a possible conformal 
diagram. Each point in this diagram represents a hyperboloid with the
 geometry of $H^3.$ The shock wave of debris emanating from the
collision disrupts the singularity. Our local brane reaches the
conformal infinity of AdS before the Cauchy horizon. It remains to be
seen whether this picture will be confirmed numerically.

\vskip 5pt

\noindent {\bf Acknowledgements:}
We thank C.~Carvalho, D.~Coule, J.~Garriga, 
S.~Ghassemi, G.~Gibbons, F.~Glanois, C.~Gordon, F.~Quevedo,
N.~Turok, and T.~Wiseman for useful discussions.
JJBP acknowledges support from the DAMTP Relativity Group's PPARC 
Rolling Grant. MB thanks Mr Dennis Avery for supporting this work.


\begin{thebibliography}{99}

\bibitem{rs}
L. Randall and R. Sundrum, ``An Alternative to Compactification,"
Phys. Rev. Lett. 83, 4690 (1999) (hep-th/9906064);
L. Randall and R. Sundrum, ``A Large Mass Hierarchy From a
Small Extra Dimension," Phys. Rev. Lett. 83, 3370 (1999) (hep-ph/9905221).

\bibitem{gt}
J. Garriga and T. Tanaka, ``Gravity in the Brane World,"
Phys. Rev. Lett. 84, 2778 (2000).
(hep-th/9911055).

\bibitem{langloisa}
P. Bin\'etruy, C. Deffayet, U. Ellwanger and  D. Langlois,
``Brane Cosmological Evolution in a Bulk With Cosmological Constant,"
Phys. Lett. B477, 285 (2000) (hep-th/9910219);
P. Bin\'etruy, C. Deffayet and D. Langlois,
``Nonconventional Cosmology From a Brane Universe,"
Nucl. Phys. B565, 269 (2000) (hep-th/9905012).

\bibitem{trodden}
G. Starkman, D. Stojkovic and M.
Trodden, ``Homogeneity, Flatness and `Large' Extra Dimensions,"
Phys. Rev. Lett. 87, 231303 (2001) (hep-th/0106143);
G. Starkman, D. Stojkovic and M. Trodden,
``Large Extra Dimensions and Cosmological Problems,"
Phys. Rev. D63, 103511 (2001)
(hep-th/0012226).

\bibitem{bucher}
M.~Bucher, ``A Braneworld Universe From Colliding Bubbles,''
Phys. Lett. B530, 1 (2002) (hep-th/0107148).

\bibitem{hawkingb}
S.~W.~Hawking, I.~G.~Moss and J.~M.~Stewart,
``Bubble Collisions in the Very Early Universe,''
Phys.\ Rev.\ D 26, 2681 (1982).

\bibitem{wu}
Z.~C.~Wu, ``Gravitational Effects in Bubble Collisions,''
Phys.\ Rev.\ D  28, 1898 (1983).

\bibitem{poisson}
E. Poisson and W. Israel,
``Internal Structure of Black Holes,"
Phys. Rev. D41, 1796 (1990).

\bibitem{ac}
S.~R.~Coleman and F.~De Luccia,
``Gravitational Effects on and of Vacuum Decay,''
Phys.\ Rev.\ D {\bf 21}, 3305 (1980);
L.~F.~Abbott and S.~R.~Coleman,
``The Collapse of an Anti-De Sitter Bubble,''
Nucl.\ Phys.\ B {\bf 259}, 170 (1985).

\bibitem{capo}
J.J. Blanco-Pillado and M. Bucher, 
``Cosmological Perturbations Generated in the Colliding
Bubble Braneworld Universe," Phys. Rev. D65, 83517 (2002) (hep-th/0111089).

\bibitem{gtb}
J. Garriga and T. Tanaka, ``Cosmological Perturbations in the 5-d Big Bang,''
Phys. Rev. D65, 103506 (2002) (hep-th/0112028).

\bibitem{capob}
J.J. Blanco-Pillado, M. Bucher, F. Glanois and S. Ghassemi, in preparation.

\end{thebibliography}
\end{document}